%% file: main.tex
\documentclass[runningheads]{llncs}
\usepackage{amsmath,amssymb,amsfonts}
\usepackage{cite}
\usepackage{algorithmic}
\usepackage{textcomp}
\usepackage{xcolor}

\usepackage[english]{babel}
\usepackage{array, booktabs, tabularx}
\usepackage{graphicx, caption, subcaption, xcolor}
\usepackage{hyperref}
\usepackage[euler]{textgreek}

\usepackage{listings}
\usepackage[T1]{fontenc}
\usepackage{lmodern}

\usepackage[title]{appendix}

\definecolor{isarblue}{HTML}{006699}
\definecolor{isargreen}{HTML}{009966}
\lstdefinelanguage{isabelle}{%
    keywords=[1]{type_synonym,datatype,fun,abbreviation,definition,proof,lemma,theorem,corollary},
    keywordstyle=[1]\bfseries\color{isarblue},
    keywords=[2]{where,assumes,shows,and},
    keywordstyle=[2]\bfseries\color{isargreen},
    keywords=[3]{if,then,else,case,of,SOME,let,in,O},
    keywordstyle=[3]\color{isarblue},
}
\newcommand{\sprod}[1]{\langle #1\rangle}

\usepackage{bbding}

\begin{document}
\title{
Verification of NP-hardness Reduction Functions for Exact Lattice Problems
\thanks{This work was supported by the Research Training Group GRK 2428 CONVEY of the
German Research Council (DFG).}}
%
%
\author{Katharina Kreuzer\orcidID{0000-0002-4621-734X}\Envelope \and\\
Tobias Nipkow\orcidID{0000-0003-0730-515X}
}
\authorrunning{K. Kreuzer and T. Nipkow}
%
\institute{
Technical University of Munich\\
Boltzmannstr. 3, 85748 Garching, Germany 
}
\maketitle              
\begin{abstract}
This paper describes the formal verification of NP-hardness reduction functions 
of two key problems relevant in algebraic lattice theory:
the closest vector problem
and the shortest vector problem, both in the infinity norm. The formalization
uncovered a number of problems with the existing proofs in the literature.
The paper describes how these problems were corrected in the formalization.
The work was carried out in the proof assistant Isabelle.

\end{abstract}
\input{CVP_SVP}

\subsubsection{Acknowledgements} 
We thank Manuel Eberl for continuous support and fruitful discussions. The first author gratefully acknowledges the financial
support of this work by the research training group ConVeY funded by the German Research Foundation under
grant GRK 2428.

%
%
%
\bibliographystyle{splncs04}
\bibliography{bibliography}

\begin{subappendices}
\renewcommand{\thesection}{\Alph{section}}%
\input{Appendix}

\end{subappendices}

\end{document}

%% file: CVP_SVP.tex
\section{Introduction}

%
%
In recent years, algebraic lattices have received increasing attention for their use in post-quantum cryptography. 
Algebraic lattices are additive, discrete subgroups of $\mathbb{R}^n$, i.e.\ a set of points in $\mathbb{R}^n$ with certain structures. One can also define lattices over finite fields, rings or modules as used in many modern post-quantum crypto systems such as the CRYSTALS suites, NTRU and Saber.
%

Two problems form the very basis for computationally hard problems on lattices, namely the closest vector problem (CVP) and the shortest vector problem (SVP). 
Given a finite set of basis vectors in $\mathbb{R}^n$, the set of all linear combinations with integer coefficients forms a lattice. In optimization form, the SVP asks for the shortest vector in the lattice and the CVP asks for the lattice vector closest to some given target vector, both with respect to some given norm. 

When working over the reals, the $p$-norm (for $p\geq 1$) is defined as $\sqrt[p]{\sum_i |x_i|^p}$. 
The most common examples are the Euclidean norm $\Vert x\Vert_2$ and the infinity norm $\Vert x\Vert_\infty = \max_i \{ |x_i| \}$, which is the limit for $p\rightarrow \infty$. 

We have formalized, corrected and verified a number of NP-hardness proofs from the literature, uncovering a number of mistakes along the way.
The first NP-hardness proof of the CVP and SVP in infinity norm is due to van Emde-Boas \cite{EmBo81}.
For other norms (especially for the Euclidean norm), there is only a randomized reduction for the NP-hardness of the SVP so far \cite{Ajt98}. For the CVP, NP-hardness has been shown in any $p$-norm for $p\geq 1$. One exemplary proof can be found in the book by Micciancio and Goldwasser~\cite[Chapter 3, Thm 3.1]{Mic02}.


The CVP and SVP were the starting point for lattice-based post-quantum cryptography \cite{Mic09}. Moreover, the relevance of these problems can also be seen from the rich literature on approximation results. For example, the LLL-algorithm by Lenstra, Lenstra and Lov\'{a}sz \cite{LLL} gives a polynomial-time algorithm for lattice basis reduction which solves integer linear programs in fixed dimensions. Using this reduced basis, one can find good approximations to the CVP using Babai's algorithm~\cite{Babai86} for certain approximation factors. 
Still, for arbitrary dimensions, the problem remains NP-hard. 
Further approximation results for the CVP, SVP and integer programming can be found elsewhere \cite{Mic98,DKRS,Khot05,HR07,roth21}. 
These approximation problems are used in cryptography.
However, we will focus on the exact CVP and SVP in this paper.

A number of more basic NP-hardness proofs have been formalized in several theorem provers so far. For example, there are formalizations of the Cook-Levin Theorem in Coq~\cite{CookLevinCoq} and Isabelle~\cite{CookLevinIsabelle}. Formalizing Karp's 21 NP-hard problems (including the Subset Sum and Partition Problems assumed to be NP-hard in this paper) in Isabelle is an ongoing project.

\subsection{Contributions}

In this paper we present NP-hardness proofs of the CVP and SVP in infinity norm that have been verified in a proof assistant.
We roughly follow the book by Micciancio and Golwasser~\cite[Chapter 3, Thm 3.1]{Mic02} and the report by van Emde-Boas~\cite{EmBo81}. However, many problems with the original proofs were encountered during the formalization efforts. We will have a look at different approaches and their advantages or problems.

We also verified the proof of NP-hardness of the CVP for any finite $p \ge 1$ from the book by Micciancio and Goldwasser.
This verification did not uncover any problems with the informal proof. Thus we do not discuss it in detail.

These formalizations were carried out with the help of the proof assistant \href{https://isabelle.in.tum.de/index.html}{Isabelle}~\cite{LNCS2283,Concrete} and are available online \cite{CVP_Hardness-AFP}. They comprise 5200 lines.
To the authors knowledge, they are the first formalizations of hardness proofs for lattice problems.
Because of the importance of the SVP and CVP and the problems in existing proofs, we consider our proofs a contribution to the foundations of verified cryptography. However, we do not claim that these hardness results directly imply quantum-resistance of any lattice-based cryptosystems.

\subsection{Overview}

The paper is structured as follows.
Section~\ref{Foundations} introduces the foundations. The rest of the paper is dedicated to the proofs, which are phrased as the following two polynomial time reduction chains:
\begin{itemize}
\item Subset Sum $\leq_p$ CVP
\item Partition $\leq_p$ Bounded Homogeneous Linear Equations $\leq_p$ SVP
\end{itemize}
Subset Sum and Partition are famous fundamental problems whose NP-hardness has been proved many times in the literature and which we take for granted.

Section~\ref{CVP} presents the reduction of Subset Sum to the CVP. Differences between our formalization and the book by Micciancio and Goldwasser~\cite{Mic02} are presented with examples that demonstrate problems with the original proof. Moreover, an example is given why the generalization to the SVP given in \cite{Mic02} does not work. 

Therefore we turn to the early proof of NP-hardness of the SVP by van Emde Boas~\cite{EmBo81}.
This proof uses the Bounded Homogeneous Linear Equations problem (BHLE) which is introduced in Section~\ref{BHLE}. 
The formalization of this proof is one of the major achievements in this paper. It posed a significant challenge since it often relied on human intuition and had to be restructured appropriately to allow a formal proof. The main proof steps are explained and difficulties in the formalization effort are described.
This proof only works in infinity norm and we explain why.
In Section~\ref{SVP}, the reduction from BHLE to the SVP is given. Again, this proof was quite elaborate to formalize as there were inaccuracies and a lot of intuition was involved. Differences between the formal proof and \cite{EmBo81} are explained by examples.

In Section~\ref{Euclidean}, we have a quick look at the reduction proof for the CVP in $p$-norm (for finite $p\ge 1$). 
In the case of the SVP there only exists a randomized hardness proof in Euclidean norm by Ajtai~\cite{Ajt96} up to now. 

Finally, the time complexity of the reduction functions are considered in Section~\ref{Time}.
We conclude the paper with a short summary and outlook.

\section{Foundations}\label{Foundations}

This section introduces known foundations mainly to fix the terminology and notation: problem reductions, lattices, and the combinatorial problems under consideration (CVP, SVP, Partition and Subset Sum).





\subsection{Problem Reductions}



Formally, a \emph{decision problem} is given by the set of \emph{YES-instances} $P$ and a set $\Gamma$ of problem \emph{instances}, where $P\subseteq \Gamma$.
We often associate the decision problem with the set of YES-instances, when the instance set $\Gamma$ is obvious and not explicitly defined.
In this paper we will often phrase problems informally (e.g.\ ``decide if $p$ is prime'') rather than give them explicitly as sets. 
For example, the decision problem ``decide if a natural number $p$ is prime'' will be formalized in the following way: the set of problem instances is $\Gamma = \mathbb{N}$ (in Isabelle these are all elements of type $nat$); and the YES-instances are $P = \{ p\in\mathbb{N} \mid p \text{ is prime} \}$ (in Isabelle this is a set of type $nat\ set$).

\begin{definition}[Problem reduction]\label{Karp-reduction}
Let $A \subseteq \Gamma$ and $B \subseteq \Delta$ be two problems.
A function $f: \Gamma\to\Delta$ is a reduction from $A$ to $B$ if it fulfills the following properties:
\begin{itemize}
  \item $\forall a\in \Gamma.\ a\in A \Leftrightarrow f(a) \in B$
  \item $f$ can be computed in polynomial time
\end{itemize}
\end{definition}
If $A$ is NP-hard, a reduction to $B$ proves NP-hardness of $B$.


In this paper we present reduction functions informally (e.g.\ ``an $a$ is reduced to a $b$ that is constructed like this'') and often with copious amounts of ``\dots'' to construct vectors etc. Of course in the formalization these reduction functions are spelled out in complete detail.
Since all operations used in the reduction functions in this paper are elementary, the polynomial time property has not been formalized but is briefly discussed in Section~\ref{Time}.
The focus of our paper are the proofs $a \in A \Leftrightarrow f(a) \in B$.

\subsection{Lattice-based Computational Problems}

To have a better understanding, we will first introduce lattices as such. Lattices are a structured set of points. They form an additive, discrete subgroup of $\mathbb{R}^n$. Formally, we define the following.
\begin{definition}[Lattice]
Let $A = \{a_1,\dots,a_n\}\subset \mathbb{R}^n$ be a set of linearly independent vectors. Then the integer span of $A$ forms a lattice $\mathcal{L}$, that is:
\begin{equation*}
\mathcal{L} = \left\{ \sum_{i=1}^n c_ia_i \mid c_i\in\mathbb{Z}\right\}
\end{equation*}
\end{definition}

Examples of lattices in $\mathbb{R}^n$ can be found in Appendix~\ref{ex:lattice}.
In the rest of the text and in the formalization we restrict to finite bases over $\mathbb{Z}$ (instead of $\mathbb{R}$), simply for computability reasons. Of course bases over $\mathbb{Q}$ can be transformed into bases over $\mathbb{Z}$ by scaling all basis vectors.


The starting point of most known hard problems on lattices are the shortest vector problem and the closest vector problem. They are defined below (as usual in decision and not in optimization form).
The lattice $\mathcal{L}\subseteq \mathbb{Z}^n$ is assumed to be generated by a finite basis in $\mathbb{Z}^n$.

\begin{definition}[Closest Vector Problem (CVP)]\label{CVP_def}
Given a lattice $\mathcal{L}$, a vector $b\in \mathbb{Z}^n$ and an estimate $k$, decide whether there exists a  vector $v\in \mathcal{L}$ such that 
\begin{equation*}
\Vert v-b\Vert \leq k
\end{equation*}
\end{definition}

\begin{definition}[Shortest Vector Problem (SVP)]\label{SVP_def}
Given a lattice $\mathcal{L}$ and an estimate $k$, determine whether there exists a vector $v\in\mathcal{L}$ such that 
\begin{equation*}
\Vert v\Vert\leq k \text{ and } v\neq 0
\end{equation*}
\end{definition}

Examples of CVP and SVP instances can be found in Appendix~\ref{ex:CVP_SVP}.

\subsection{Partition and Subset Sum Problems}

Recall that we plan to prove NP-hardness of the CVP and SVP in the case of the infinity norm by reducing the well-studied NP-complete Subset Sum and Partition problems to the CVP and SVP.
We state the definitions.

\begin{definition}[Partition problem]
Given a finite list of integers $a_1,\dots,a_n$, does there exist a partition of $\{1\dots n\}$ into subsets $I$ and $\{1\dots n\}\setminus I$ such that
\begin{equation*}
\sum_{i\in I} a_i = \sum_{i \in \{1\dots n\}\setminus I} a_i
\end{equation*}
\end{definition}



The Partition problem can be seen as a special case of the Subset Sum problem.

\begin{definition}[Subset Sum problem]
Given a finite list of integers $a_1,\dots, a_n$ and an integer $s$, decide whether there exists a subset $S$ of $\{1\dots n\}$ such that 
\begin{equation*}
\sum_{i\in S} a_i = s
\end{equation*}
\end{definition}


\subsection{Notation}

Throughout the paper we use traditional mathematical notation,
in particular the graphical ``$...$''. The formal Isabelle notation is by necessity more verbose (and precise).
Our formalization employs both lists and vectors as a type for finite sequences and converts between them where necessary.
For reasons of presentation we blur this distinction in the paper.

\section{CVP}\label{CVP}

In this section, we formalize the proof of the NP-hardness of the CVP in the infinity norm along the lines of \cite[p 48., Chapter 3.2, Thm 3.1]{Mic02} by reducing Subset Sum to the CVP.

An instance $a_1,\dots,a_n,s$ of Subset Sum
is mapped to the following instance of the CVP:
\begin{equation}\label{eq:red_cvp2}
\mathcal{L} = 
\begin{pmatrix}
a_1 &\cdots &a_n\\
a_1 &\cdots &a_n\\
2 & & 0\\
&\ddots & \\
0& & 2\\
\end{pmatrix}
\cdot \mathbb{Z}^{n}
\qquad
b = 
\begin{pmatrix}
s-1\\
s+1\\
1\\
\vdots \\
1\\
\end{pmatrix}
\qquad k=1
\end{equation}

We proved the following theorem:
\begin{theorem}\label{thm:NP_CVP}
The above mapping is a reduction from the Subset Sum problem to the CVP (in infinity norm).
\end{theorem}
This implies that the CVP (in infinity norm) is an NP-hard problem.

The reduction function used by Micciancio and Goldwasser~\cite{Mic02} actually looks a bit different. The image of $a_1,\dots,a_n,s$ would be 
\begin{equation}\label{eq:red_cvp1}
B = \begin{pmatrix}
a_1 &\cdots &a_n\\
2 & & 0\\
&\ddots & \\
0& & 2\\
\end{pmatrix}
\qquad
\mathcal{L} = 
B\cdot \mathbb{Z}^{n}
\qquad
b = 
\begin{pmatrix}
s\\
1\\
\vdots \\
1\\
\end{pmatrix}
\qquad k=1
\end{equation}

However, the proof in \cite[p.49]{Mic02} with this reduction function works only for $p<\infty$. It goes along the lines of the following idea:
Take $k = \sqrt[p]{n}$. In the case of $p = \infty$, we get 
$
k = \lim_{p\rightarrow \infty} \sqrt[p]{n} = 1
$.
Then we can formulate the following equality (equation (3.5) in \cite[p.49]{Mic02}):
\begin{equation}\label{eq:Bxt}
\Vert Bx-b\Vert ^p_p = \left| \sum _{i=1}^n a_ix_i-s \right| ^p + \sum_{i=1}^n |2x_i-1|^p
\end{equation}

Given a YES-instance $a_1,\dots,a_n,s$ of Subset Sum, there exists a vector $x = (x_1,\dots,x_n) \in \{0,1\}^n$, such that $ \sum _{i=1}^n a_ix_i-s  = 0$ and $|2x_i-1| = 1$. Then $\Vert Bx-b\Vert ^p_p = n$ which proves this case.

Given a YES-instance of the CVP defined by $\mathcal{L}$, $t$ and $k$ that are the image of $a_1,\dots,a_n,s$ under the reduction function as in \eqref{eq:red_cvp1}, we get $\Vert Bx-b\Vert^p_p\leq n$. Since all values are integers, we have $|2x_i-1| \geq 1$. It follows that $\sum _{i=1}^n a_ix_i-s  = 0$ and $|2x_i-1| = 1$. Thus, we can deduce that $a_1,\dots, a_n, s$ was indeed a YES-instance of Subset Sum.

The major problem we encountered was that this proof works fine for $p<\infty$ but for $p=\infty$, the sum in \eqref{eq:Bxt} becomes a maximum instead. 
The equation then reads
\begin{align*}
\begin{split}
\Vert Bx -  b\Vert_\infty = \max \left(\left| \sum _{i=1}^n a_ix_i-s \right|,  |2x_i-1| \text{ for } 1\leq i \leq n\right)
\end{split}
\end{align*}
This invalidates the arguments in the proof since $\left| \sum _{i=1}^n a_ix_i-s \right|$ can now be in the range $\{-1,0,1\}$. The constraints are too lax to ensure the equality to zero. 

A solution was to alter the matrix and target vector and add another entry. The matrix and target vector we used are 
given in equation~\eqref{eq:red_cvp2}.
The alternation to $s-1$ and $s+1$ forces a linear combination of the $a_i$ to be exactly $s$ in the hardness proof, since 
$
|\sum_i c_i a_i - (s\pm 1)|\leq 1
$.

After communicating with Daniele Micciancio, one of the authors of \cite{Mic02}, he suggested using a constant $c > 1$ and the generating instance
\begin{equation*}
\mathcal{L} = 
\begin{pmatrix}
c\cdot a_1 &\cdots & c\cdot a_n\\
2 & & 0\\
&\ddots & \\
0& & 2\\
\end{pmatrix}
\cdot \mathbb{Z}^{n}
\qquad
b = 
\begin{pmatrix}
c\cdot s\\
1\\
\vdots \\
1\\
\end{pmatrix}
\qquad k=1
\end{equation*}
This solves the problem as well and can be implemented using e.g.\ $c=2$. This technique is described later in the book \cite[p.49-51]{Mic02} when trying to explain the NP-hardness proof for the SVP in the infinity norm.

\subsection{Towards the SVP}

The authors of \cite{Mic02} argue that the reduction argument of the SVP can be deduced generating an instance of the SVP using the Subset Sum instance $a_1,\dots,a_n,s$ in the following way. For $c>1$, e.g.\ $c=2$, take
\begin{equation*}
B = 
\begin{pmatrix}
c\cdot a_1 &\cdots & c\cdot a_n & c\cdot s\\
2 & & 0 & 1\\
&\ddots & & 1\\
0& & 2 & 1\\
\end{pmatrix} 
\qquad
\mathcal{L} =
B\cdot \mathbb{Z}^{n+1}
\qquad k=1
\end{equation*}
The authors claim that every shortest vector in the image of the reduction function has $-1$ as last coefficient. For example, let a YES-instance of the SVP be defined by the generating matrix $B$ of the lattice
and let $x= (x_1,\dots,x_n,-1)^T$ be the coefficients such that $B x$ is a shortest vector.
Then we know that 
\begin{equation*}
\Vert Bx\Vert_\infty = 
\left| \left|
\begin{pmatrix}
c\cdot (x_1 a_1 +\dots + x_n a_n - s)\\
2x_1 - 1\\
\vdots\\
2x_n - 1\\
\end{pmatrix}
\right| \right|_\infty \le 1
\end{equation*}
Since $c>1$, it follows, that $x_1 a_1 +\dots + x_n a_n - s = 0$, which yields a solution for the given Subset Sum instance $a_1,\dots,a_n,s$.


However, this reduction does not always work as the following example shows:

\begin{example}
Given the Subset Sum instance 
$
(a_1,a_2,a_3,s) = (1,1,1,1)
$.
This is a YES-instance, since a solution is given by 
$x_1=1$, $x_2=0$ and $x_3=0$.
The basis matrix of the corresponding SVP would be (with $c>1$) 
\begin{equation*}
B =
\begin{pmatrix}
c& c& c& c\\
2& 0& 0& 1\\
0& 2& 0& 1\\
0& 0& 2& 1\\
\end{pmatrix}
\end{equation*}

Take for example the vector $v = B\cdot (-1,-1,-1,3)^T = (0,1,1,1)^T$. 
It has infinity norm $1$ and is thus a shortest vector in the lattice generated by $B$.
However, this vector has the last coefficient $3$ and not $-1$, even though it clearly is a shortest vector of the lattice given by $B$.
The corresponding scaled ``solution'' for Subset Sum would be $(1/3,1/3,1/3,-1)$ but since only integer values are allowed in the solution space, this is not a solution in our sense.

We consider another example. Let the Subset Sum instance be $a_1' = 3, s' = 1$. We can easily see that this is not a YES-instance, i.e.\ there exists no solution. Still, the corresponding SVP instance given via the reduction function is generated by the matrix 
\begin{equation*}
B' =
\begin{pmatrix}
c\cdot 3& c\cdot 1\\
2& 1\\
\end{pmatrix}
\end{equation*}
In this case the coefficients $(-1,3)^T$ yield a shortest vector in the lattice spanned by $B'$, since
\begin{equation*}
\left| \left| B'
\begin{pmatrix}
-1 \\ 
3\\
\end{pmatrix}\right| \right| _\infty = 
\left| \left|
\begin{pmatrix}
0\\
1\\
\end{pmatrix}
\right| \right|_\infty \le 1
\end{equation*}
Thus, $B'$ defines a YES-instance of the SVP, but the original Subset Sum instance is not a YES-instance.


In \cite{Mic02}, it is stated for the infinity norm that any shortest vector yields a solution for the Subset Sum Problem, which is not the case in these examples:  we cannot ensure that a shortest vector always has $-1$ as a last coordinate.
\end{example}

Although the proof in \cite{Mic02} does not work out as expected, there is still the reduction proof by van Emde-Boas \cite{EmBo81}
which reduces a problem called the Bounded Homogeneous Linear Equation problem to the SVP in infinity norm. This will be discussed in the next two sections.

\section{Bounded Homogeneous Linear Equations}\label{BHLE}
A technical report by Peter van Emde-Boas \cite{EmBo81} gives another reduction proof for the NP-hardness of the SVP in infinity norm.
The author first reduces the Partition Problem to a problem called Bounded Homogeneous Linear Equation (BHLE) which is then reduced to the SVP.
\begin{definition}[Bounded Homogeneous Linear Equations problem]\\
Given a finite vector of integers $b \in\mathbb{Z}^n$ and a positive integer $k$, decide whether there exists an $x\in \mathbb{Z}^n\setminus \{0\}$ with $\Vert x\Vert_\infty \leq k$ such that
\begin{equation*}
\sprod{b,x} = 0
\end{equation*}
\end{definition}

We have verified a reduction from Partition to BHLE, and thus BHLE is NP-hard.
\begin{theorem}
There is a reduction from Partition to BHLE in infinity norm.
\end{theorem}
The proof is carefully engineered and rather intricate. Differences to the original proof and problems encountered during the formalization are:

\begin{itemize}
\item Our formal proof has a different structure than the proof in the technical report \cite{EmBo81}. 
Indeed, the technical report first proves the reduction of a weaker form of Partition to BHLE and then argues that ``omitting'' an element yields the desired result as it adds stricter constraints. In the formalization we skip this intermediate step and directly prove the existence of an appropriate reduction function.
\item Steps that seem trivial in the technical report often require a long formal proof. What can be reasoned by intuition in a pen-and-paper proof has to be elaborated in the formal proof. Intuition is also sometimes used for hand-waving over small gaps or imprecisions.
\item Indexing vectors and lists has been a problem in the formalization. In pen-and-paper proofs, one can argue easily about ``omitting'' an element of a list even though this is imprecise and often misuses the notation. In the formalization one cannot simply skip an index. All indexing functions in the formalization have to be total. ``Omitting'' an element can only be solved by re-indexing and re-structuring the lists in the proof.
\item Numbers are interpreted in different number systems during the proof. 
In contrast to the original proof, the formalization has to explicitly state the digits for a change of basis and show equivalence. This leads to verbose and elaborate proofs.
To make proofs easier, we use the concrete basis $d=5$ instead of an unspecified basis $d>4$ as in \cite{EmBo81}.
Furthermore, the number $M$ must use the absolute values of the $a_i$ (omission in the definition of $M$ in \cite{EmBo81}). The formal definition is stated below.
\item The proof involved many arguments about manipulations of huge sums. 
Working with huge sums entails very large proof states where the existing proof automation mostly failed on.
These proof states require detailed (but still readable) proofs and occasional manual instantiation of theorems.
Another possible solution to get smaller proof states is to introduce local abbreviations for subterms.
\end{itemize}

Let us have a look at the proof and its difficulties in the formalization in more detail.
We start from a Partition instance $a = a_1,\dots, a_n$ . Note that we ignore the trivial case $n=0$ in this presentation (but deal with it in the formal proofs) --- this means $n-1 \ge 0$. 
We reduce $a$ to a BHLE instance $b$ as follows:
\begin{itemize}
  \item Define
  	\begin{equation}\label{eq:M}
  		M = 2\cdot(\sum_{i=1}^n |a_i|) + 1
  	\end{equation}
  \item For $1\leq i < n$ generate a 5-tuple
  \begin{align}
    b_{i,1} &= a_i + M \cdot (5^{4i-4} + 5^{4i-3} + 5^{4i-1}) \label{eq:b_start}\\
    b_{i,2} &=       M \cdot (5^{4i-3} + 5^{4i}) \nonumber\\
    	b_{i,3} &=       M \cdot (5^{4i-4} + 5^{4i-2}) \nonumber\\
    	b_{i,4} &= a_i + M \cdot (5^{4i-2} + 5^{4i-1} + 5^{4i}) \nonumber\\
    	b_{i,5} &=       M \cdot (5^{4i-1}) \nonumber\\
    	b_i &= b_{i,1},b_{i,2},b_{i,4},b_{i,5},b_{i,3}\nonumber
  \end{align}
  Note that $b_{i,3}$ has moved to the last position in $b_i$.
  \item For $i=n$ generate only a $4$-tuple:
  \begin{align}
    b_{n,1} &= a_n + M \cdot (5^{4n-4} + 5^{4n-3} + 5^{4n-1}) \nonumber\\
    b_{n,2} &=       M \cdot (5^{4n-3} + \textcolor{red}{1}) \nonumber\\
    	b_{n,4} &= a_n + M \cdot (5^{4n-2} + 5^{4n-1} + \textcolor{red}{1}) \nonumber\\
    	b_{n,5} &=       M \cdot (5^{4n-1})\label{eq:b_end}\\
    	b_n &= b_{n,1},b_{n,2},b_{n,4},b_{n,5}\nonumber
  \end{align}
  Note that
  \begin{itemize}
      \item $b_{n,3}$ is omitted from $b_n$ to restrict the constraints necessary for the proof and
      \item that in $b_{n,2}$ and $b_{n,4}$ the last summand changes to a $+1$ in comparison to the other $b_{i,2}$ and $b_{i,4}$.
  \end{itemize}
\end{itemize}
In summary, the entry $b_{i,3}$ is uniformly in the last position in the $b_i$ but omitted from the final $b_n$.

The Partition instance $a$ of length $n$ is reduced to a vector $b$ of length $5n-1$: 
\begin{equation}\label{eq:b}
    b = (b_1,\dots,b_{n-1},b_n)
\end{equation}

The NP-hardness proof now follows in three steps:
\begin{enumerate}
\item We need to show an auxiliary lemma.
\item We show that a YES-instance of Partition is reduced to a YES-instance of BHLE.
\item We show that the pre-image of a YES-instance of BHLE is indeed a YES-instance in Partition.
\end{enumerate}

\subsection{Auxiliary Lemma}
As a first step, the proof needs a short auxiliary lemma from number theory. 
\begin{lemma}\label{lemma_help}
Let $x, y, c \in \mathbb{Z}^n$ and $M$ be an integer.
Assume that $M > \sum_{i=1}^n |x_i|$ and that $|c_i|\leq 1$ for all $1 \le i \le n$. Furthermore, let the following equation hold:
\begin{equation}\label{lemma_assms}
\sum_{i=1}^n c_i \cdot (x_i + M \cdot y_i) = 0
\end{equation} 
Then we have 
\begin{equation*}
\sprod{c, x} = 0\quad\text{ and }\quad \sprod{c, y} = 0
\end{equation*}
\end{lemma}
In this lemma, we can reinterpret $x_i + M \cdot y_i$ from \eqref{lemma_assms} as a number in basis $M$ with lowest digit $x_i$. Even with a coefficient $c_i$, the lowest digit in basis $M$ has to be zero, as well as the rest. By splitting off the lowest digits consecutively, we can show, that indeed all digits in basis $M$ have to equal zero.

\subsection{$a \in$ Partition $\Longrightarrow$ $b \in$ BHLE}
This direction is quite easy. 
Let $a_1,\dots,a_n$ be a YES-instance of partition with partitioning set $I$. 
We will show that the following vector $x$ is a solution to the corresponding BHLE:
\begin{align*}
    x &= (x_1,\dots,x_{n-1},x_n)\\
    x_i &= \begin{cases} 
    	1,-1,0,-1,0 & i\in I \land n-1 \in I\\ 
    	0,0,-1,1,1 &  i\in I \land n-1\notin I\\
    0,0,-1,1,1 & i\notin I \land n-1 \in I\\
    1,-1,0,-1,0 & i\notin I \land n-1\notin I
    	\end{cases}\qquad 1 \le i < n\\
    x_n &= 1,-1,0,-1\\
\end{align*}

We have to show that $\langle b, x\rangle = 0$.
This is proven by plugging in the definitions and rearranging terms in the sum of the scalar product such that they cancel out.
As a last step in the proof, we need to show that $\Vert x\Vert_\infty \leq 1$. For the infinity norm this is quite easy. 
However, it would not be true for other norms. For $p\geq 1$ and $p<\infty$ we have for $n\geq 1$:
\begin{equation*}
\Vert x\Vert_p = \sqrt[p]{3n}>1
\end{equation*} 
Thus, the chosen constraints $x$ only work in infinity norm. 
The explicit proof can be found in the Appendix~\ref{proof:BHLE1}.

\subsection{$a \in$ Partition $\Longleftarrow$ $b \in$ BHLE}
This direction is harder. 
Let $b$ be a YES-instance of BHLE. That is, there exists a nonzero $x$ such that 
$\sprod{b,x} = 0$ and $\Vert x\Vert_\infty \leq 1$. 
We have to show that there is a partition $I$ on $a_1,\dots, a_n$ with $\sum_{i\in I} a_i = \sum_{i\in \{1\dots n\}\backslash I} a_i$.

The proof idea works as follows. First, we apply the auxiliary lemma and get a constraint on the $a_i$ on the one hand, and a condition on the $x_i$ with coefficients that are powers of $5$ on the other hand.
Using this condition on the $x_i$, we generate equational constraints on the entries of $x$ by looking at the digits in basis 5. 
We argue that a number equals zero if and only if all its digits are zero.

The generated equations lead to a good characterisation of $x$, namely the weight $w = x_{5(n-1)+1}$.
From the assumption that $\Vert x\Vert_\infty \leq 1$, we deduce $|w|\leq 1$.
Again, this step can only be reasoned in the infinity norm. For other $p$-norms, this argumentation breaks as we need the property $|w|\leq 1$ to complete the proof.
Using the value of $w$, we can constuct a partitioning set $I$ with the required property from the equation on the $a_i$.
The explicit proof can be found in Appendix~\ref{proof:BHLE2}.


\section{SVP}\label{SVP}
Knowing that the BHLE is indeed an NP-hard problem, we reduce it to the SVP. Then we can conclude that the SVP in infinity norm is NP-hard.

\begin{theorem}\label{thm:SVP_NP}
There is a reduction from BHLE to the SVP in infinity norm. 
\end{theorem}

Again some difficulties were met when formalizing the proof for the above theorem.
First of all, note that the terminology in \cite{EmBo81}
and nowadays is a bit different. In \cite{EmBo81}, the shortest vector problem only denotes the shortest vector problem in the Euclidean norm. What we call the shortest vector problem in the infinity norm is named closest vector problem in \cite{EmBo81}.
To make terminology even more confusing, our understanding of the closest vector problem is called the nearest vector problem in \cite{EmBo81}.
To make the notation clear, we provide a table for reference in the Appendix~\ref{app:notation}, Figure~\ref{tab1}.

A more mathematical problem encountered was that the reduction itself used in \cite{EmBo81} was not entirely correct.
In the reduction two factors $k'=k+1$ and $k''$ were introduced. These factors should have certain properties to allow the arguments of the reduction proof to go through. However, this is only true when tweaking these factors a bit to make the whole proof watertight. We will now have a closer look.

Given the BHLE instance $b = (b_1,\dots, b_n)$ and $k$, create the following SVP instance:
\begin{equation*}\label{eq:redSVP}
\mathcal{L} = 
\begin{pmatrix}
1 & & 0 & 0\\
 & \ddots &  & \vdots\\
0 & & 1 & 0\\
-&(k+1)\cdot b & - & k''\\
\end{pmatrix}
\cdot \mathbb{Z}^{n}
\qquad
k=k
\end{equation*}
where $k''$ is the factor in question. In the technical report, we have 
$$k'' = 2\cdot (k+1)\cdot (\sum_i b_i) +1$$
The following example however shows that this factor is not enough.

\begin{example}
Consider the BHLE instance given by \mbox{$b=(1,-1)$} and $k=1$.
This is a YES-instance, since the vector $(1,1)$ yields the expected properties.

Define the following matrices.
\begin{equation*}
B_0 = 
\begin{pmatrix}
1 & 0 & 0\\
0 & 1 & 0\\
2 & -2 & 1\\
\end{pmatrix}
\quad\quad
B_1 = 
\begin{pmatrix}
1 & 0 & 0\\
0 & 1 & 0\\
2 & -2 & 9\\
\end{pmatrix}
\quad\quad
B_2 = 
\begin{pmatrix}
1 & 0 & 0\\
0 & 1 & 0\\
6 & -6 & 25\\
\end{pmatrix}
\end{equation*}

The associated SVP instance is the lattice generated by $B_0$.
Then the vector $(0,0,1)^T$ with infinity norm $1$ is a solution to the SVP instance generated by the basis matrix $B_0$. However, since the last entry is nonzero, this does not provide a solution for BHLE. Contrary to this example, the proof in the technical report shows that for all SVP solutions the last entry must be zero.

The reason, why the argument in the technical report breaks at this point is because $b_1 + b_2 = 0$, thus making $k'' = 1$ very small. One step to prevent this is to use the absolute values of the $b_i$ in $k''$ instead.
The new $k''_1$ we consider is 
\begin{equation*}
k''_1 = 2\cdot (k+1)\cdot (\sum_i |b_i|) +1
\end{equation*}

With this new factor $k''_1$ we get the generating matrix $B_1$
and the vector $(0,0,1)$ is no longer a shortest vector.

Still, this is not enough. Consider the same $b=(1,-1)$ as above, but let $k=5$.
Then we get $B_2$ as the generating matrix of the SVP lattice.
The vector $x=(0,5,1)^T$ is a shortest vector whose last entry is nonzero. Again it contradicts the proof in the technical report.
The reason this time is the following: the argument that \mbox{$(k+1)\left(\sum_{i=1}^n x_i b_i\right)$} and $k''_1$ have different relative sizes fails. 
Indeed, we have
\begin{equation*}
\left|\left| 
\begin{pmatrix}
1 & 0 & 0\\
0 & 1 & 0\\
6 & -6 & 25\\
\end{pmatrix}
\cdot
\begin{pmatrix}
0\\
5\\
1\\
\end{pmatrix}
\right|\right|_\infty = 
\left|\left|\begin{pmatrix}
0\\5\\-5\\
\end{pmatrix}
\right|\right|_\infty = 
5 \leq k
\end{equation*}

\end{example}

We can obtain different relative sizes of $(k+1)\left(\sum_{i=1}^n x_ib_i\right)$ and $k''_1$ by defining 
\begin{equation}\label{k'_def}
k''_2 = 2\cdot \textcolor{red}{k} \cdot (k+1)\cdot (\sum_i |b_i|) +1
\end{equation}

Now we can make sure that the last entry of a solution to the SVP problem is indeed zero. 
For the proof of Theorem~\ref{thm:SVP_NP} we consider the reduction given by 
\begin{equation*}
\mathcal{L} = 
\underbrace{
\begin{pmatrix}
1 & & 0 & 0\\
 & \ddots &  & \vdots\\
0 & & 1 & 0\\
-&(k+1)\cdot b & - & k''_2\\
\end{pmatrix}
}_{B}
\cdot \mathbb{Z}^{n}
\qquad
k=k
\end{equation*}
where $B$ denotes the basis matrix generating the lattice $\mathcal{L}$ as given above.

Consider a solution \mbox{$x = (x_1,\dots,x_{n+1})$} of the SVP with $\Vert Bx\Vert_\infty \leq k$. 
Then we have  
\begin{align*}
\begin{split}
Bx= &
\begin{pmatrix}
1 & & 0 & 0\\
 & \ddots &  & \vdots\\
0 & & 1 & 0\\
-&(k+1)\cdot b & - & k''_2\\
\end{pmatrix}
\cdot 
\begin{pmatrix}
x_1\\
\vdots\\
x_n\\
x_{n+1}\\
\end{pmatrix} 
= 
\begin{pmatrix}
x_1\\
\vdots\\
x_{n}\\
(k+1)(\sum_{i=1}^n x_i b_i) + x_{n+1} \cdot k''_2\\
\end{pmatrix} 
\end{split}
\end{align*}
As this yields a solution to the SVP, we get:
\begin{equation}\label{eq:contradiction}
|(k+1)(\sum_{i=1}^n x_i b_i) + x_{n+1} \cdot k''_2|\leq k
\end{equation}
Then we calculate:
\begin{align*}
\begin{split}
(k+1)(\sum_{i=1}^n x_i b_i) + x_{n+1} \cdot k''_2 
&\leq (k+1)(\sum_{i=1}^n |x_i| |b_i|) + x_{n+1} \cdot k''_2 \leq \\
&\leq (k+1)k(\sum_{i=1}^n |b_i|) + x_{n+1} \cdot k''_2 
\end{split}
\end{align*}

Assuming that $x_{n+1}\neq 0$, we have 
\begin{align*}
\begin{split}
|(k+1)k(\sum_{i=1}^n |b_i|)| &< |2\cdot k \cdot (k+1)\cdot (\sum_i |b_i|) +1 | 
= |k''_2| \leq |x_{n+1} \cdot k''_2|
\end{split}
\end{align*}
Thus the two summands indeed have different relative sizes and can never cancel out the other summand. This leads to a contradiction to \eqref{eq:contradiction}. Therefore, $x_{n+1}=0$ must be true and $(x_1,\dots,x_n)$ constitutes a solution to the BHLE when using $k''_2$ as in \eqref{k'_def}.

\section{Other $p$-Norms}\label{Euclidean}
Up to now, we have investigated lattice problems under the infinity norm. Even though this yields nice hardness results, in practice the Euclidean norm is used more often. Unfortunately, when considering $p$-norms things do not play out as nicely. In this section, we assume $1\le p<\infty$ whenever we talk about a specific $p$.

For the CVP, there is a generalisation of the proof for every $p$-norm in \cite[p.48, Chapter 3.2, Thm 3.1]{Mic02} which we also formalized.
Let $a_1,\dots,a_n,s$ be an instance of Subset Sum. The reduction function maps this instance to:
\[
\mathcal{L} = 
\begin{pmatrix}
a_1 &\cdots &a_n\\
2 & & 0\\
&\ddots & \\
0& & 2\\
\end{pmatrix}
\cdot \mathbb{Z}^{n}
\qquad
b = 
\begin{pmatrix}
s\\
1\\
\vdots \\
1\\
\end{pmatrix}
\qquad k=\sqrt[p]{n}
\]

Then the following theorem holds:
\begin{theorem}\label{thm:NP_CVP_p}
The above mapping is a reduction from the Subset Sum problem to the CVP in $p$-norm.
\end{theorem}
This implies that the CVP in $p$-norm is an NP-hard problem.
The outline to the proof is given in Section~\ref{CVP} after Theorem~\ref{thm:NP_CVP}. The important difference to the infinity norm is that the bound $k$ scales with the dimension $n$ of the lattice. 

%

For the SVP, there is no known deterministic NP-hardness result in the Euclidean norm, or even any $p$-norm. 
However, Ajtai~\cite{Ajt96,Ajt98} found an interesting alternative which is quite useful for the application in cryptography, namely randomized reductions using polynomial-time probabilistic reduction functions.
In cryptography, these results guarantee the hardness of ``average'' cases. That is, given an average instance according to a probability distribution, it will most likely be intractable.

%
%

\section{Time complexity}\label{Time}
As stated in Section~\ref{Foundations}, time complexity of the above reduction functions has not been formalized. However, we give a short explanation why all reduction functions are indeed in polynomial time.


\textbf{Subset Sum to CVP:}
The reduction function as given in equation \eqref{eq:red_cvp2} creates $(n+2)(n+1)+1$ values using only memory access or one addition. 
Therefore, the time complexity in this case is $\mathcal{O}(n^2)$.

\textbf{Partition to BHLE:}
In this case, the reduction function maps the input $a$ of length $n$ to $b$ as defined in equation~\eqref{eq:b}.
The value $k=1$ is fixed. Then $a$ is mapped to a vector of length $5n-1$. 
When calculating the $b_i$, we need to calculate the value of $M$ as in \eqref{eq:M}. As we sum over all input values, this lies in $\mathcal{O}(n)$.
Each $b_i$ can then be calculated in $\mathcal{O}(n)$ since it only contains a constant number of additions of the input with fixed cofactors (see \eqref{eq:b_start} - \eqref{eq:b_end}). 
Putting the construction of the list and the calculation of the $b_i$ together, we find that the whole reduction function is in $\mathcal{O}(n^2)$.

\textbf{BHLE to the SVP:}
Consider the reduction function as given in equation~\eqref{eq:redSVP} using the value $k''_2$ as in \eqref{k'_def}.
Calculating $k''_2$ requires $n+2$ memory accesses which are processed in $n+4$ arithmetic operations, thus having a time complexity of $\mathcal{O}(n)$.
Every other entry in the matrix is calculated on $\mathcal{O}(1)$, since they contain at most two memory accesses and at most two arithmetic operations.
The input generates $(n+1)^2+1$ values, of which $(n+1)(n+1)$ are in $\mathcal{O}(1)$ (namely all the zeros and ones, the vector $(k+1)\cdot a$ and the constraint $k$) and one is calculated in $\mathcal{O}(n)$ (namely $k''_2$). 
Thus, the whole reduction function lies in $\mathcal{O}(n^2)$.

\section{Outlook}
With this paper, we now have a formal proof for NP-hardness of the CVP and SVP in the infinity norm, as well as a formal proof of the CVP in $p$-norm (for $1\le p <\infty$). In the formalization process, many gaps and imprecisions in the pen-and-paper proofs were fixed. The changes to the original proofs have been elaborated with explanations and examples.
Unfortunately, giving a deterministic reduction proof of the SVP in $p$ norm for $p<\infty$ is still an open problem. Under probabilistic assumptions, Ajtai showed NP-hardness of the SVP in Euclidean norm in \cite{Ajt98}. 

An interesting topic for future work is to develop 
a framework
for probabilistic reductions such as in \cite{Ajt98}. 
This will give the foundation to extend formalization of hardness proofs to other problems in lattice theory, especially those used in lattice-based cryptography, such as the Learning with Errors (LWE) Problem, Ring-LWE and Module-LWE. 
This will underline the security of many lattice-based crypto systems. 
Another topic for future work is to formalize the hardness proofs for approximate versions of the CVP and SVP.

%% file: Appendix.tex
\section{Examples of Lattices}\label{ex:lattice}
\begin{example}
\begin{figure}[htb]
\centering
\begin{subfigure}{0.5\textwidth}
	\centering
	\includegraphics[width=0.725\textwidth]{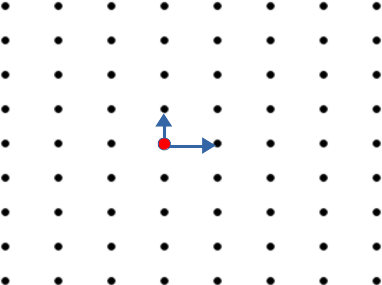}
	\caption{Lattice with rectangular basis vectors}
	\label{fig:rect_lattice}
\end{subfigure}%
\begin{subfigure}{0.5\textwidth}
	\centering
	\includegraphics[width=0.6\textwidth]{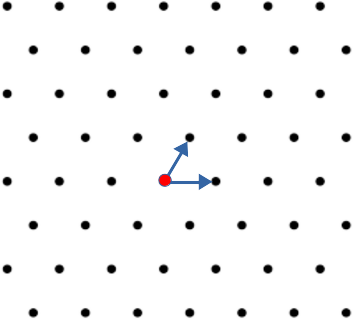}
	\caption{Lattice with triangular basis vectors}
	\label{fig:triang_lattice}
\end{subfigure}
\caption{Two exemplary lattices in $\mathbb{R}^2$}
\label{fig:lattice}
\end{figure}
In Figure~\ref{fig:lattice} two examples of lattices in $\mathbb{R}^2$ are depicted. The red point is the origin. The two blue arrows show the basis vectors $a_1$ and $a_2$ that are linearly independent and span the lattice. Every integer combination of the two blue arrows is a black point, an element of the lattice. 

We can see that the grid spanned by the basis vectors is discrete and has some recurring structures. These structures are determined by the basis vectors: the angle between them and their length. In Figure~\ref{fig:rect_lattice}, the angle between the two basis vectors is $90^{\circ}$ yielding a rectangular fundamental domain. Whereas in Figure~\ref{fig:triang_lattice}, we have an angle of $60^{\circ}$ between the basis vectors and equal length. This produces a fundamental domain of an equilateral triangle. 

Indeed, the automorphism group of a lattice is a symmetry group, see Conway~\cite[Chapter 3.4]{Con99}. For example, in Figure~\ref{fig:rect_lattice} the symmetry group is \textbf{pmm} and in Figure~\ref{fig:triang_lattice} is it \textbf{p3m1}\cite{Liu98}.
\end{example}

\section{Examples of instances of the CVP and SVP}\label{ex:CVP_SVP}
\begin{figure}
\centering
\begin{minipage}{.5\textwidth}
  \centering
  \includegraphics[width = 0.9\textwidth]{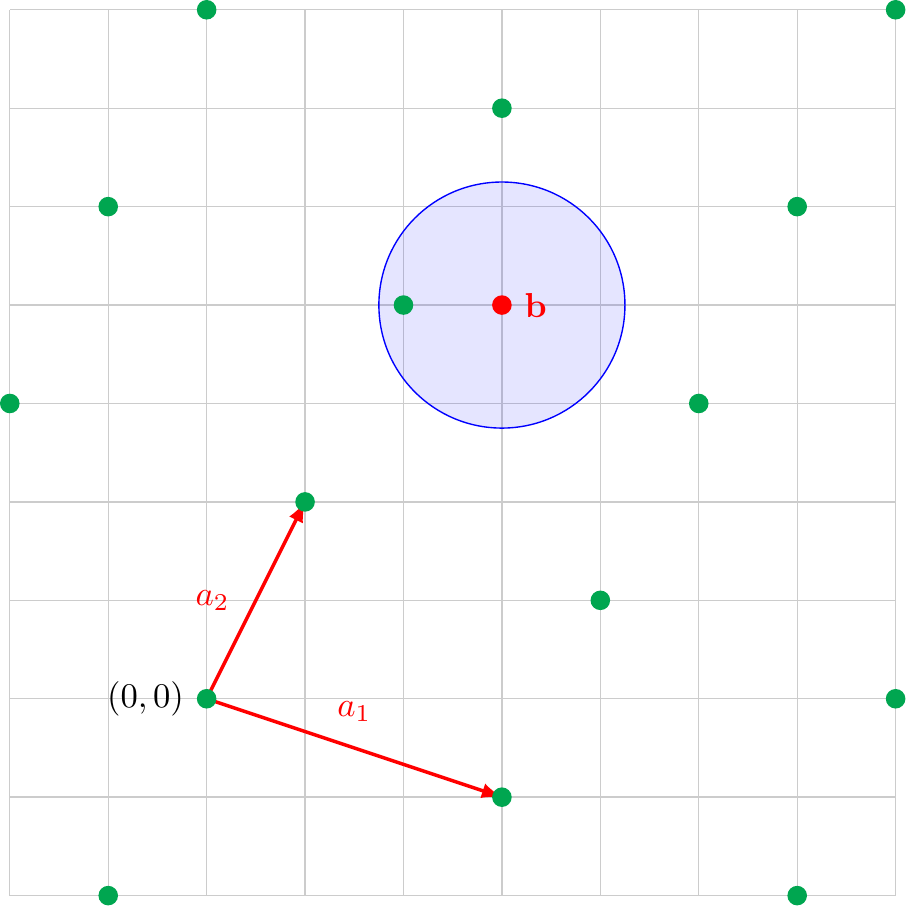}
  \captionof{figure}{An instance of the CVP in $\mathbb{Z}^2$}
  \label{fig:CVP}
\end{minipage}%
\begin{minipage}{.5\textwidth}
  \centering
  \includegraphics[width = 0.9\textwidth]{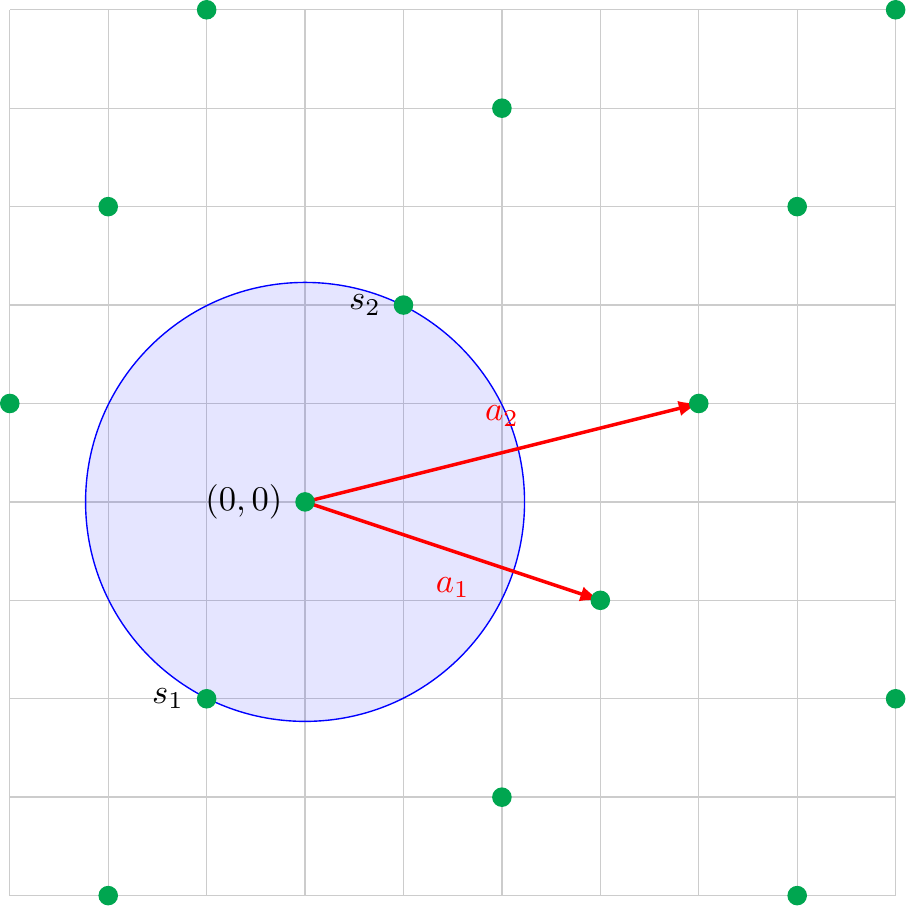}
  \captionof{figure}{An instance of the SVP in $\mathbb{Z}^2$}
  \label{fig:SVP}
\end{minipage}
\end{figure}

\begin{example}
Figure~\ref{fig:CVP} shows a two-dimensional instance of the CVP in Euclidean norm. The green points form the lattice $\mathcal{L}$ which is spanned by the two red vectors $a_1$ and $a_2$. The target vector is the red point labeled $b$. The estimate $k$ is depicted as the radius of the blue circle around $b$. 

In this case, we have a YES-instance, since there exists a lattice point close enough to the target vector (there is a green point in the blue circle around $b$). Indeed, the green dot in the blue circle is a solution point to the search problem associated to the CVP.
\end{example}

\begin{example}
In Figure~\ref{fig:SVP}, an instance of the SVP in $\mathbb{Z}^2$ in Euclidean norm is depicted. The lattice $\mathcal{L}$ is drawn as the set of green points. It is generated by the two red vectors $a_1$ and $a_2$. The estimate $k$ is the radius of the blue ball around the origin (annotated by $(0,0)$). 

In this case, we have a YES-instance of the SVP. There are two points, namely $s_1$ and $s_2$ which are on the edge of the blue circle around the origin. As there are no other green points inside the blue circle apart from the origin, therefore $s_1$ and $s_2$ are indeed the shortest vectors of the lattice. 

This is a nice example to see that there always exist at least two shortest vectors. The reason is very simple: Assume there is a shortest vector $v$, then also $-v$ is a shortest vector, since $\Vert v\Vert = \Vert -v\Vert$ in any norm. In our case, $s_1$ and $s_2$ both are possible solutions to the search problem of the SVP.
\end{example}

\section{Proofs for BHLE}\label{proof:BHLE}
\subsection{Proof of ``$a \in$ Partition $\Longrightarrow$ $b \in$ BHLE''}\label{proof:BHLE1}
\begin{proof}
Let $a_1,\dots,a_n$ be a YES-instance of Partition with partitioning set $I$. We will show that the following vector $x$ is a solution to the corresponding BHLE:
\begin{align*}
    x &= (x_1,\dots,x_{n-1},x_n)\\
    x_i &= \begin{cases} pm & i\in I\\
    mp & \text{otherwise} \end{cases}\qquad 1 \le i < n\\
    x_n &= 1,-1,0,-1\\
    pm &= \begin{cases} 1,-1,0,-1,0 & n-1 \in I\\ 0,0,-1,1,1 & \text{otherwise} \end{cases}\\
    mp &= \begin{cases} 0,0,-1,1,1 & n-1 \in I\\
    1,-1,0,-1,0 & \text{otherwise} \end{cases}
\end{align*}


We can now calculate the following: 
\begin{align*}
\sprod{b,x} &= \sum_{i=1}^{5n-1} b_i \cdot x_i = \\
&= (\sum_{i=1}^{5(n-1)} b_i \cdot x_i ) + \sprod{(b_{n,1},b_{n,2},b_{n,4},b_{n,5}), (1,-1,0,-1)} = \\
&= (\sum_{i=1}^{n-1} \sprod{(b_{i,1},b_{i,2},b_{i,4},b_{i,5},b_{i,3}), 
(\mathit{if}\ i\in I\ \mathit{then}\ pm\ \mathit{else}\ mp)} )\quad + \\
&\qquad + b_{n,1}-b_{n,2}-b_{n,5} = \\
&= (\sum_{i\in I\cap \{1...n-1\}} b_{i,1}-b_{i,2}-b_{i,5}) + (\sum_{i\in \{1\dots n-1\}\setminus I} -b_{i,4}+b_{i,5}+b_{i,3})\quad +\\
&\qquad + b_{n,1}-b_{n,2}-b_{n,5} = \\
&= (\sum_{i\in I\cap \{1...n-1\}} a_{i} + M \cdot (5^{4i-4} - 5^{4i}))\quad +\\
&\qquad + (\sum_{i\in \{1...n-1\}\setminus I} - a_i + M \cdot (5^{4i-4} - 5^{4i})) + a_n + M \cdot (5^{4n-4} - 1 ) = \\
&= (\sum_{i\in I} a_{i})\quad  -\quad (\sum_{i\in \{1...n\}\setminus I} a_i) \quad + \\
&\qquad	+ M\cdot \left(5^{4n-4} - 1 + \sum_{i\in\{1...n-1\}} 5^{4i-4} - 5^{4i} \right) =\\
&= 0\\
\end{align*}
For the last equality, we need two facts: 
Firstly, since  $a$ is a YES-instance of Partition with partitioning set $I$, we have 
\begin{equation*}
\sum_{i\in I} a_{i} = \sum_{i\in \{1\dots n\}\setminus I} a_{i}
\end{equation*}
Secondly, $M$ is multiplied by a telescopic sum that reduces to zero.

As the entries of $x$ are in $\{-1,0,1\}$, we have $\Vert	x\Vert_\infty \leq 1$.
All in all, $x$ constitutes a solution for the BHLE instance given by $b$ and $1$.
\end{proof}

\subsection{Proof of ``$a \in$ Partition $\Longleftarrow$ $b \in$ BHLE''}\label{proof:BHLE2}
\begin{proof}
Let $b$ be a YES-instance of BHLE. That is, there exists an $x$ such that 
$\sprod{b,x} = 0$ and $\Vert x\Vert_\infty \leq 1$. Again, this step only works for the infinity norm. We will look more closely at this later in the proof.

The proof goal is to find a set $I\subseteq \{1\dots n\}$ such that 
\begin{equation*}
\sum_{i\in I} a_i = \sum_{i\in \{1\dots n\}\setminus I} a_i
\end{equation*}

Unfortunately, we do not know the exact values of $x$, so we need to derive equational constraints on $x$.
We have
\begin{align*}
\begin{split}
0 &= \sprod{b,x} =  \sum_{i\in \{1\dots 5n-1\} } b_i \cdot x_i  =\\
&= \left(\sum_{i\in\{0\dots n-1\}} (x_{5i + 1} + x_{5i+3}) \cdot a_{i+1}\right)  + \  M \cdot \left( \sum_{i\in\{1\dots 5n-1\}}x_i \cdot c_i \right) 
\end{split}
\end{align*}
where $c = (c_1,\dots, c_{5n-1})$ is the appropriate rest consisting only of sums over powers of $5$. 

We observe that $M$ was chosen in a manner such that 
\begin{equation}\label{eq:sum_less_M}
\left|\sum_{i\in\{0\dots n-1\}} (x_{5i+1} + x_{5i+3}) \cdot a_{i+1} \right| < M
\end{equation}
From Lemma~\ref{lemma_help} we know that each digit has to be zero if the whole number equals to zero if the assumptions hold. Therefore, knowing \eqref{eq:sum_less_M} and $|x_i|\leq 1$ by the assumptions that $b$ is a YES-instance of BHLE, the following equations are derived immediately. 
\begin{equation}\label{eq0}
\sum_{i\in\{0\dots n-1\}} (x_{5i+1} + x_{5i+3}) \cdot a_{i+1} = 0
\end{equation}
\begin{equation}\label{eq0_rest}
\sum_{i\in\{1\dots 5n-1\}}x_i \cdot c_i = 0
\end{equation}

Since every summand in \eqref{eq0_rest} consists of a power of $5$ times an element of $x$, we can rewrite this sum as a number in basis $5$ by accumulating all coefficients to a power of $5$. We denote the digits by a function $d$ of the index.
\begin{equation*}
0 = \sum_{i\in\{1\dots 5n-1\}}x_i \cdot c_i = \sum_{k\in\{0\dots 4n-1\}}d(k) \cdot 5^{k}
\end{equation*}

Again, applying Lemma~\ref{lemma_help} consecutively with $|d(k)| <5$ (we split off the lowest digit in the representation in basis $5$) yields that every digit $d(k)$ equals zero for $k<4n$.
This yields the following equations:
\begin{align*}
\begin{split}
\forall i\in & \{1\dots n-1\} :\\
& x_{5i+1} + (\mathit{if\ }i<n-1\mathit{\ then\ }x_{5i+5}\mathit{\ else\ }0)+ x_{5(i-1)+2} + x_{5(i-1)+3} = 0\\
\end{split}
\end{align*} 
\begin{align*}
\begin{split}
& x_1 + (\mathit{if\ }1<n\mathit{\ then\ }x_5\mathit{\ else\ }0) +  x_{5(n-1)+2} + x_{5(n-1)+3} = 0\\
\end{split}
\end{align*}
\begin{align*}
\forall i\in\{ 0\dots n-1\} :\quad x_{5i+1} + x_{5i+2} = 0
\end{align*}
\begin{align*}
\begin{split}
\forall i\in & \{ 0\dots n-1 \} :\quad(\mathit{if\ }i<n-1\mathit{\ then\ }x_{5i+5}\mathit{\ else\ }0) + x_{5i+3} = 0
\end{split}
\end{align*}
\begin{align*}
\forall i\in\{ 0\dots n-1\} :\quad x_{5i+1} + x_{5i+3} + x_{5i+4} = 0
\end{align*}

From these equations, we can derive that the value $x_{5i+1} + x_{5i+5}$ does not depend on $i$ (for $i<n-1$). We call this value the weight $w$ where
\begin{equation*}
w = x_{5i+1} + x_{5i+5}
\end{equation*}
The observant reader notices that the definition of the weight is the reason we needed to omit the last element in the vector $b$. Indeed, the element $x_{5(n-1)+5}$ is not defined and for $i=n-1$ the weight is only
\begin{equation*}
w = x_{5(n-1)+1}
\end{equation*}
This constrains the bound on the absolute value $|w|\leq 1$, since $\Vert x\Vert_\infty \leq 1$. 

It is essential to constrain the weight to $|w|\leq 1$, since otherwise we cannot deduce a partition. Assume $|w| = 2$, then also a solution with $x_{5i+1} + x_{5i+3} = x_{5i+1} - x_{5i+5} = 0$, i.e.\ $x_{5i+1} = x_{5i+5} = 1$ is allowed. Then, \eqref{eq0} does not yield a partition as it is an empty sum.

Since we work over the integers, we only need to consider the values $w\in\{-1,0,1\}$.
Here, the solution $w=0$ leads to $x=0$, a contradiction to the assumptions that $x$ is a solution to the BHLE instance $b$.
Thus, we will only look at the case of $w=1$. The case $w=-1$ proceeds analogous with flipped signs.

Using the above equations, we can conclude that either $x_{5i+1}=1 \wedge x_{5i+5}=0$ or $x_{5i+1}=0 \wedge x_{5i+5}=1$.
Then, \eqref{eq0} yields the desired partition for the YES-instance of the Partition Problem.
This concludes the proof.
\end{proof}

\section{Different notations}\label{app:notation}

\begin{figure}
\begin{center}
\begin{tabular}{c|c}
technical report \cite{EmBo81}
& our notation\\
\hline
closest vector problem & SVP in infinity norm\\
shortest vector problem & SVP in Euclidean norm\\
nearest vector problem & CVP\\
\end{tabular}
\end{center}
\caption{Notation \label{tab1}}
\end{figure}